\documentstyle[12pt]{article}

\begin{document}
\title{ Towards a New Strategy  \\
   of Searching for QCD Phase Transition \\
  in Heavy Ion Collisions }
\author { M.~P\l{}oszajczak$^{ \dagger}$, A.A.~Shanenko $^{\ddagger}$
and V.D.~Toneev$^{ \dagger , \ddagger} $ \vspace*{2mm} \\
\small \em  $^{\dagger} $
Grand Acc\'{e}l\'{e}rateur National d'Ions Lourds (GANIL),
BP 5027, \\
\small \em F-14021 Caen Cedex, France \\
\small \em $^{ \ddagger}$  Bogoliubov Laboratory of Theoretical Physics,
Joint Institute for Nuclear Research,\\
\small \em 141980 Dubna, Moscow Region,  Russia\\ }
\date{}
\maketitle

\begin{abstract}
We reconsider the  Hung and Shuryak  arguments in favour of searching for the
deconfinement phase transition in heavy ion collisions {\em downward} from
the nominal SPS energy, at  $E_{lab} \approx 30 \ GeV/A$ where the
fireball lifetime is the longest one.
Using the recent lattice QCD data and the mixed phase
model, we show that the deconfinement transition might occur at
the bombarding energies as low as $E_{lab}=3 - 5 \ GeV/A$. Attention is
drawn to the study of the mixed phase of nuclear matter
in the collision energy range $E_{lab}= 2-10 \ GeV/A$.
\end{abstract}

\vspace*{0.2cm}



\newpage

Over the last ten years, a fundamental QCD prediction  of the
phase transition from hadrons into a state of free quarks and gluons
(quark-gluon plasma) has been studied actively.
 Extensive lattice QCD calculations not only allowed to specify
 the deconfinement temperature, the order of phase transition and
its flavor dependence both for pure gluon matter and for the plasma with
dynamical quarks, but also gave insight into physical phenomena above
and below the deconfinement temperature. By now, there is a rather long
list of various signatures which can signal  the quark-gluon plasma
formation. Many of these signatures were tested experimentally
at CERN-SPS with the $^{16}O$ and $^{32}S$  beams at $ 200 \ GeV/A$ where
crucial conditions for the deconfinement transition are expected to be
reached. Indeed, some predicted effects such as the strangeness enhancement,
$J/\Psi$ suppression or $\phi /(\rho +\omega )$ enhancement
have been observed but their hadronic interpretation cannot be
excluded. General belief is that for more definite conclusions
{\em heavier ions} and {\em higher beam energies} should be used.

Recently, a new strategy of experimental search for the QCD
phase transition in heavy ion collisions has been advocated \cite{HS94}.
 The equation of state (EOS) is very 'soft' in a narrow range of
temperatures ($\Delta T \simeq 10 \ MeV$) around the transition temperature,
leading to a significant reduction in {\em transverse} expansion of
the fireball formed in heavy ion collisions \cite{SZ79}.  This 'softness'
of the EOS, estimated by Hung and Shuryak at the energy density
$\varepsilon_{sp} \simeq 1.5 \ GeV/fm^3$ and the beam energy
$E_{lab} \approx 30 \ GeV/A$ \cite{HS94},  affects not only the transverse
but also the {\em longitudinal } expansion and results in a longest
lifetime of the excited system. Radical changes in the hydrodynamic
space-time evolution  around the 'softest point' of the EOS should lead
for certain observables to a  sharp and specific dependence
on the {\em heavy-ion beam energy }. Therefore, some aspect of
the deconfinement transition might be better studied at lower
collision energy, {\em downward } from the nominal SPS energy.

Hung and Shuryak  are confident that even in more sophisticated
models than used in ref.\cite{HS94}, the total lifetime of a fireball
should have a maximum near the indicated collision energy region.
{}From our point of view, this conclusion is not evident because of
the two crucial assumptions made. Firstly,  the crossover
 behaviour of the deconfinement transition  was simulated
in ref.\cite{HS94}   by an {\em arbitrary smoothing}
of the results of a simple two-phase model exhibiting the first
order phase transition.  It is not clear how well this description
approximates the lattice QCD results. Secondly, the EOS of the
{\em baryon-free matter}, $n_B=0$, has been used  \cite{HS94}.
Due to a considerable stopping  power, there is little
hope to create a baryonless matter in nucleus-nucleus collisions
up to $E_{lab} \simeq 200 \ GeV/A$ (especially for heavy systems).
In this paper, we shall consider how the longest-lived fireball is
sensitive to the details of the EOS  and how it survives in baryon-rich
matter inherent in  systems formed at energies $E_{lab} \le 200 \ GeV/A$.
Our analysis is based on the statistical model taking properly
into account the mixed phase in which unbound quarks and gluons coexist
with hadrons \cite{SYY93a,SYY95}.

The EOS is the key quantity and thermodynamic properties of excited
matter near QCD phase transition should be calculated from the first
principles in a non-perturbative manner. At present, such calculations
are possible only in the lattice QCD. This approach shows that
gluonic matter in pure gauge $SU(2)$ and $SU(3)$ theories exhibits
a phase transition of the second and first order, respectively
(for a recent review see ref.\cite{U95}).
For the more realistic case of the $SU(3)$ theory with dynamical quarks,
recent lattice studies \cite{U95} show that there is a smooth crossover
rather than a distinct phase transition if the quark masses are close to
the physical ones.  For the most important case of the lattice QCD
at finite baryon number density, nothing is really known and
statistical models should be invoked to describe thermodynamic
properties of the excited nuclear matter. There are many versions of the
statistical model (see ref.\cite{CGS86} and references quoted therein)
but all of them predict the deconfinement phase transition  of the
first order and, therefore, do not reproduce the lattice QCD results
with dynamical quarks mentioned above. The only exception
is the statistical mixed phase model of the deconfinement
 \cite{SYY93a,SYY95}, which we will use in our considerations.
The application of the mixed phase models to the deconfinement
transition is supported strongly by the discussion of deconfinement as
a color screening effect and the observation that the hadron-like
excitations survive above the critical temperature
 (see for example ref.\cite{SS95}).

The specific feature of the approach  developed by Shanenko, Yukalova and
Yukalov (SYY) \cite{SYY93a,SYY95} is to consider the coexistence
of spatially non-separated the hadron phase and the quark-gluon
plasma phase.The latter one consists of unbound 'generic' particles
(quarks and gluons in our case) while hadron phase is composed
of quark-gluon clusters. In the mixed phase model \cite{SYY93a},
one assumes beforehand the separation of cluster degrees of freedom,
i.e.  an exact Hamiltonian $H(\psi)$ is replaced by an effective
cluster Hamiltonian  $H_{c} (\psi_c )$.
Here, $\psi$ denotes field operators of the generic particles and
$ \psi_c \equiv \{ \psi_n;n=1,2,\ldots \} $ stands for quasiparticle
operators ($n>1$ corresponds to clusters and $n=1$  to unbound generic
particles). A large variety of quasiparticles leads to an enormous
number of possible states. In the  equilibrium, the physical state
of a system corresponds  to the extremum of the thermodynamic potential
$F(H_c)$.

The following two important points should be emphasized. Firstly,
the effective Hamiltonian $H_c(\psi_c )$ may acquire extra dependence on
thermodynamic parameters like temperature $T$ and cluster densities
$\rho_c$: $ H_c \equiv H_c (\psi_c,T,\rho_c ) $ where
$\rho_c \equiv \{\rho_n;n=1,2,\ldots \}$. The appearance
of a density-dependent interaction is a distinctive feature
of the SYY approach as compared to  other statistical models \cite{CGS86}
\footnote{Similar situation is met in the case when interaction
is taken into account by the excluded-volume method.}. Secondly,
 one must ensure that $H(\psi )$ and $H_c(\psi_c )$ are thermodynamically
equivalent, i.e.  their thermodynamic characteristics are the same in
the thermodynamic limit of $V \to \infty $ keeping constant $\rho_c$
\cite{Y91}. These demands of the thermodynamic equivalence
and the thermodynamic self-consistency impose additional conditions:
\begin{eqnarray*}
 \lim_{V \to \infty \atop \rho_c =const }
  \ \frac{1}{V} \left[ \ F(H) - F(H_c) \ \right]  = 0 \ ,
\end{eqnarray*}
\begin{eqnarray}
 \lim_{V \to \infty \atop \rho_c =const }
  \ \frac{1}{V} \left[ \ dF(H) -dF(H_c) \ \right]  = 0
\label{eq2a}
\end{eqnarray}
which lead to
\begin{eqnarray}
 \left< \frac{\partial H_c}{\partial T} \right> &=& 0 \ , \label{eq3a} \\
 \nonumber  \\
\left< \frac{\partial H_c}{\partial \rho_n} \right> &=& 0,
\ \ \ \ \ \ n=1,2,\ldots
\label{eq3}
\end{eqnarray}
and essentially define the form of the cluster Hamiltonian. In the
mean-field approximation, the cluster Hamiltonian
becomes \cite{SYY93a}:
\begin{eqnarray}
 H_c = \sum_n \sum_s \int d{\bf r} \ \psi^+_n({\bf r},s) \
\left( \ K_n+U_n(T,\rho_c) \ \right) \
\psi_n({\bf r},s) - C \cdot V \ ,
\label{eq4}
\end{eqnarray}
where $n$ enumerates the clusters and $s$ stands for their internal degrees
of freedom. $K_n=\sqrt{-\nabla^2 + M^2_n}$ is the kinetic energy and
$U_n$ is a mean field acting on the $n$-particle cluster.
In the same approximation, the condition
of thermodynamic equivalence (\ref{eq3a}) becomes:
\begin{eqnarray}
 \frac{\partial U_n(T,\rho_c)}{\partial T} = 0 \ \ \ \ \  n=1,2,\ldots ,
\label{eq5}
\end{eqnarray}
i.e. the mean fields $U_n=U_n(\rho_c )$ depend on the temperature only
through the temperature dependence of the  densities $\rho_n$.
The $c$-number term $C\cdot V$  in eq.(\ref{eq4}) is necessary to satisfy
the conditions (\ref{eq3a}) and (\ref{eq3}) in which case $C$ is
a function of $U_n$'s~\cite{Y91}.
Unbound particles are treated as trivial clusters with $n=1$.

If masses $M_n$ of the isolated clusters and their quantum numbers are known
either experimentally or from other calculations, then, to apply the mixed
phase model, one has to define only $U_n$. The mean field
$U_1$ acting on unbound gluons or quarks can be approximated as
follows~\cite{SYY93a,SYY95}:
\begin{eqnarray}
  U_1 = \frac{A}{\rho^{\gamma}} \  \ \ \ \ \ \ \ \ \ \ 0 < \gamma < 1
\label{eq6}
\end{eqnarray}
where  $\rho = \sum_n n \ \rho_n$ is the total quark-gluon density.
The presence of $\rho$ in (\ref{eq6})
corresponds to the inclusion of the interaction
between all components of the mixed phase. Note that when there are
no hadrons, i.e.  $\rho_n=0$ for $n>1$, the expression (\ref{eq6})
is just the same as used earlier to describe the thermodynamic
properties of the quark-gluon plasma with a density-dependent quark
mass  \cite{B85}. For the case of $n>1$, alongside with
the mean-field term  (\ref{eq6})
depending on the constants $A$ and $\gamma$, there is also a
cluster-cluster interaction potential which is  proportional to the hadron
density and the function $\Phi_{nm}(r)$ characterizing
the interaction strength between clusters of $n$ and $m$ generic
particles \cite{SYY93a,SYY95}. For long-ranged cluster interactions
it is possible to get a recurrent relation \cite{Y91}:
\begin{eqnarray}
  \Phi_{nm} (r) \sim nm \ \Phi_0(r) \ ,
\label{eq7}
\end{eqnarray}
and to reduce all unknown interactions to the single interaction potential
$\Phi_0(r)$ between simplest non-trivial clusters, e.g. two-gluon
glueballs in the ground state,  lightest mesons or  baryons.
In the Hartree approximation to $H_c$,  one needs to know
only one constant: $ {\bar{\Phi}}_0  \equiv \int d{\bf r} \ \Phi_0(r)$
to describe the cluster-cluster interactions. Thus,
the Hamiltonian (\ref{eq4}) is completely defined  and
thereby any thermodynamic characteristics of the mixed phase system
can be found if three parameters $A$, $\gamma$ and ${\bar{\Phi}}_0$
have been fixed \cite{SYY93a,SYY95}. This 3-parameter set was
found \cite{SYY93a} by fitting a temperature dependence of the energy
density $\varepsilon$ and the pressure $p$ calculated within the
lattice QCD for the pure gluonic matter in the gauge $SU(2)$
\cite{EFR89} and $SU(3)$\cite{K89} theories. It is worth
emphasizing that  $A$ depends only on the colour group
and $\gamma$ is constant for all  gauge systems. So, fitting the
$SU(3)$ pure gluonic QCD lattice data allows  to fix $A$ and $\gamma$
parameters which can then be used for the $SU(3)$ system with quarks. As to
${\bar{\Phi}}_0$ in this case, it can be found from a nucleon-nucleon
potential by referring to the relation (\ref{eq7}). In the following,
we use $\gamma =0.62$, $A^{1/(3\gamma +1)} = 225 \ MeV $ and
${\bar{\Phi}}_0 = 4.1\cdot 10^{-5} \ MeV^{-2}$.

The mixed phase model predictions for the $SU(3)$ system with two
light flavours and $n_B=0$ are shown in Fig.~1.
When confronted with the lattice QCD data, the mixed phase model
\cite{SYY93a} gives a very similar temperature dependence of
the energy density $\varepsilon$ and pressure $p$. As follows from these
results, the mixed phase model estimates the deconfinement temperature
to be $T_{dec}=150 \ MeV$ and predicts the crossover-type phase transition
\cite{SYY93a,SYY95} in full agreement with the QCD lattice data.
Some overshooting of the lattice data at $T>1.5 \ T_{dec}$ is related to
neglecting the negative Coulomb-like term of the quark-gluon interactions.
Two different sets of the best available lattice QCD results
plotted in Fig.~1, correspond to two different scheme of including
quarks. A peak-like structure of $\varepsilon$ near
the deconfinement temperature $T_{dec}$ for the Kogut-Susskind
scheme \cite{B94} seems to be unrealistic, vanishing for temperatures
below  $\sim 0.9 \ T_{dec}$. The quark mass in the Kogut-Susskind
calculations amounts to $m_q \approx 0.1 \ T_{dec}$ which is close to
the physical mass used in the mixed phase model. In the Wilson scheme
we do not really know the quark mass used. In this case, the value
$m_q \sim T_{dec}$ given in \cite{CES86} seems to be enormously large
because for  $T<T_{dec}$ there is a good agreement between the Wilson
scheme results and the ideal meson gas calculations \cite{CES86}.

The EOS in the form advocated by Hung and Shuryak \cite{HS94}
is represented in Fig.2. Here, all data of Fig.1 are replotted
alongside with the curve used in \cite{HS94} to simulate the crossover
transition. The $p/\varepsilon$-functions for {\em lattice data
themselves} show {\em a minimum} which is just associated with the softest
point of the EOS where a fireball of the excited nuclear matter lives
longest. In the $p/\varepsilon$-representation, the lattice QCD data
for two schemes of accounting for dynamical quarks are strikingly different
but, nevertheless, they yield the same  position of the minimum at
$\varepsilon_{sp} \approx 0.5-0.6 \ GeV/fm^3$
which can be reached with the $Au$ beam energies of about $3-5 \ GeV/A$.
It is noteworthy that this value od $\varepsilon_{sp}$  corresponds roughly
to the value of the bag constant, so the fireball near the softest
point may be considered as a 'big hadron'. The mixed phase model predicts
the position of a minimum near $\varepsilon_{sp} \approx 0.3 \ GeV/fm^3$
which is rather close to the lattice results.  On the contrary,
 Hung and Shuryak obtain much higher value,
$\varepsilon_{sp} \approx 1.5 \ GeV/fm^3$ \cite{HS94} and
  the minimum is seen more distinctly than in
the lattice data. Since $\varepsilon \sim T^4$, the difference in
 deconfinement temperatures is small.
The Kogut-Susskind and Wilson schemes predict the deconfinement
temperatures $T_{dec}$ as high as $157 \ MeV$ \cite{B94}
and $150 \ MeV$ \cite{U95} respectively, and the mixed phase model
gives $T_{dec}=150 \ MeV$. The  Hung-Shuryak approximation does not
come from a fit to the lattice QCD data but is an arbitrary smoothing
of the two-phase model results  where $T_{dec}=160 \ MeV$ is defined
by the choosen bag constant \cite{HS94}.

The mixed phase model can be naturally generalized to the case of
the non-zero baryon density $n_B$ \cite{SYY95}. The $n_B$-dependent
pressure in the SYY approach coincides with that of the Walecka-like
model up to $n_B \approx 3 \ n_0$ \cite{HBB95}.
At higher baryon densities, the mixed phase model gives
lower values of pressure due to the quark admixture.
As to the predicted energy density, the difference between these two models
is even smaller.

As is seen in Fig.3, the $p/\varepsilon $-function changes drastically
with increasing the baryon density of a system: the position of
$\varepsilon_{sp}$ remains unchanged at $n_B < n_0$ though the
minimum is gradually vanishing with increasing $n_B$ and disappears
at $n_B \approx n_0$ . So, we arrive at somewhat controversial demands:
to reach the condition of the longest-lived fireball in heavy ion
collisions one should go {\em downward} in bombarding energies
as far as $E_{lab} \approx 3-5 \ GeV/A$ but, on the other hand,
high baryon density of a fireball formed at these energies will
suppress much the effect of the softest point. Due to  smallness of
 $\varepsilon_{sp}$ it may be more favourable to look  for the softest
point effect in non-central high-energy collisions in the target
fragmentation region. To see how the lifetime
of a fireball  will change with the beam energy, one needs
detailed dynamical calculations with EOS of the mixed phase model
 \footnote{These calculations within a hydrodynamic model are now
in progress.}. It is of interest to note that this EOS is quite
different from the EOS for the pure hadronic phase as illustrated
in Fig.3 in the case of an ideal pion gas.

Thus, the lattice QCD data do really predict a minimum
in the $p/\varepsilon$-representation of the EOS whose position,
according to Hung and Shuryak \cite{HS94}, defines the beam energy at
which the fireball formed has the longest lifetime. This representation
is significantly more sensitive, both to details of the lattice
calculations and approximations involved therein than the conventional
thermodynamic quantities $\varepsilon (T)$ and $p(T)$. The simulation
of the EOS with a crossover \cite{HS94} results in the softest point at
$\varepsilon_{sp} \approx 1.5 \ GeV/fm^3$, which is noticeably higher
than the lattice value $\varepsilon_{sp} \approx 0.5 \ GeV/fm^3$.
Describing correctly the order and temperature of the deconfinement
QCD transition for $n_B=0$, the
mixed phase model \cite{SYY93a,SYY95} predicts
$\varepsilon_{sp} \approx 0.3 \ GeV/fm^3$  and the agreement
with the lattice data may be further improved by a more accurate
treatment of cluster-cluster interactions. All this implies that
the proposed beam-energy tuning  for identification of the
deconfinement transition should be done at bombarding energies of
$E_{lab} \approx 3-5 \ GeV/A$,  much below the value
$E_{lab} \simeq 30 \ GeV/A$ advocated by Hung and Shuryak \cite{HS94}.
In the recent paper \cite{MO95}, heavy ion collisions were considered
within relativistic dissipative hydrodynamics with EOS taken from the
lattice QCD calculations  for baryonless matter. A strong enhancement of
the width of the shock front has been found  at
$E_{lab} \simeq 6 \ GeV/A$  as a manifestation of the softest point.
This value of the beam energy is in a fine agreement with our estimate,
if one  remembers that the old QCD data used in ref.\cite{MO95}
correspond to higher deconfinement temperature $T_{dec} = 200 \ MeV$.
Note that these hydrodynamic calculations are complementary
to the Hung and Shuryak estimate and show that the softest point effect
influences not only the disassembly stage but also the compression stage.

The mixed phase model predicts also a strong dependence of the EOS on
the baryon density of the system: a minimum of the $p/\varepsilon $
function is washed out for $n_B \approx n_0$. Since the state with
$\varepsilon = \varepsilon_{sp}$ is a transitional one,
we expect that the change of the fireball lifetime
with $E_{lab}$ will not be as large as predicted in \cite{HS94}.
 Nevertheless, we would like to draw attention to heavy ion
collisions at moderate ($E_{lab}=2-10 \ GeV/A$) energies for studying
the mixed phase of quarks and hadrons. As has been shown above, a pure
hadronic EOS is quite different from the EOS predicted by the mixed
phase model near the softest point. A possibility of forming the mixed
quark-hadron state at energies $E_{lab}\approx 2-10 \ GeV/A$
has been noted previously \cite{G90}. However, being based on
the two-phase models, this consideration predicts a sharp decrease of
temperature just above the deconfinement  threshold  resulting in the
formation of a 'cold' plasma. Generally speaking, the known signals of
quark-gluon plasma are not applicable to the cold plasma case.
In contrast, the mixed phase model has no threshold and no temperature
fall-off, thus such signals  should persist but their strength will be
proportional to an unbound quark abundance. It is worth mentioning that
some enhancement of the $\Lambda$-hyperon production as a specific
plasma formation signature has been observed at the energy as low  as
$3.5 \ GeV/A$ \cite{O95}. Interferometry measurements deserve
a special attention  since they are sensitive to 'granularity' of an
emitting source \cite{P94}. In this respect,
the correlation length or screening length of unbound quarks
provides a new  length scale additional to the source
radius.

\vspace*{5mm}
We are indebted to  W.~Bauer, J.~Cleymans, M.~Gorenstein, E.~Okonov, S.~Pratt,
F.~Webber and G.~Zinoviev for stimulating discussions and comments. V.D.T.
acknowledges the warm hospitality of the theory group of GANIL, Caen where
a part of this work has been done. The work of V.D.T. was supported by
Grant $N^0_-$ 3405 from INTAS (International Association for promotion
of cooperation with scientists from the independent states of the
former Soviet Union).

\newpage
{\bf {\large Figure captions} } \\

Fig.1.~Temperature dependence of energy density $\varepsilon$ and pressure
$p$ (relative to corresponding values in the Stefan-Boltzmann limit) for
$SU(3)$ gauge theory of baryon-free matter with massive dynamical quarks.
Curves are calculated within the mixed phase model \cite{SYY93a}.
Triangles and squares are lattice data for Wilson \cite{CES86}
and Kogut-Susskind \cite{B94} schemes of accounting for dynamical quarks,
respectively.

Fig.2.~The ratio of pressure and energy density $p/\varepsilon$ versus
$\varepsilon$. Notation is the same as in Fig.1. The dashed curve
corresponds to the approximation used in  ref.\cite{HS94}. Data are
given for $\varepsilon > 0.005 \ GeV fm^{-3}$.

Fig.3.~$p/\varepsilon$-representation of the EOS for the baryonic matter
predicted by the mixed phase model \cite{SYY95}. Numbers near the curves
show the baryon density in units of the normal nuclear density. The case
of ideal pion gas is given by the dashed line.


\newpage
\begin{thebibliography}{99}
\bibitem{HS94}
	C.M.~Hung and  E.V.~Shuryak,
       Preprint SUNY-NTG-94-59 (1994), to be published in Phys. Rev. Lett.
\bibitem{SZ79}
   E.~Shuryak and O.V.~Zhirov, Phys. Lett. B 89 (1979) 253; \\
   L.~Van~Hove, Z. Phys.  C 21 (1983) 93; \\ K.~Kajantie, L.~Mc~Lerran
     and P.V.~Ruuskanen, Phys. Rev. D 34 (1986) 2746; \\
   S.~Chakrabarty at al., Phys. Rev. D 46
       (1992) 3802.
\bibitem{SYY93a}
   A.A.~Shanenko, E.P.~Yukalova and V.I.~Yukalov,
   Nuovo Cim.  A 106 (1993) 1269;
   Physics of Atomic Nuclei   56 (1993) 372.
\bibitem{SYY95}
   A.A.~Shanenko, E.P.~Yukalova and V.I.~Yukalov,
   JINR Rapid Communications  1 [69]  (1995) 19;
   Physics of Atomic Nuclei 58 (1995) 335.
\bibitem{U95}
   A.~Ukawa, Report UTHEP-302 (1995).
\bibitem{CGS86}
  J.~Cleymans, R.V.~Gavai and E.~Suhonen, Phys. Rep. 130 (1986) 217.
\bibitem{SS95}
  T.~Sch\"{a}fer and E.V.~Shuryak, Phys. Lett.  B 356 (1995) 147.
\bibitem{Y91}
   A.A.~Shanenko, E.P.~Yukalova and V.I.~Yukalov,
   Physica  A 197 (1993) 629.
\bibitem{EFR89}
  J.~Engels et al., Z. Phys.  C 42 (1989) 341.
\bibitem{K89}
  F.~Karsch, Preprint CERN-TH-5498/89 (1989).
\bibitem{B85}
K.A.~Olive, Nucl. Phys. B 190 (1981) 483; {\em ibid} B 198 (1982)
 461; \\ D.H.~Boal, J.~Schachter and R.M.~Woloshin, Phys. Rev. D 26
 (1982)  3245; \\ M.~Plumer, S.~Raha and R.M.~Weiner,
Nucl. Phys.  A 418 (1984) 549c;  \\
D.~Blaschke et al., Phys. Lett.  B 151 (1985) 439; \\
I.V.~Moskalenko and D.E.~Kharzeev, Soviet Journal of Atomic Physics
 48 (1988) 713.
\bibitem{CES86}
T.~Celik, J.~Engels and H.~Satz, Nucl. Phys.  B 256 (1986) 670.
\bibitem{B94}
T.~Blum et al., Preprint AZPH-TH/94-22 (1994).
\bibitem{HBB95}
 G.~Hejc, W.~Bentz and H.~Beier, Nucl. Phys.  B 582 (1995) 401.
\bibitem{G90}
T.~Biro and J.~Zymanyi, Nucl. Phys. A 395 (1983) 25; \\
H.~St\"oker, Nucl. Phys. A 417 (1984) 587;  \\
N.~Glendenning, Nucl. Phys.  A 512 (1990) 737.
\bibitem{MO95}
 L.~Mornas and U.~Ornik,  Nucl. Phys. B 587 (1995) 828.
\bibitem{O95}
 E.O.~Okonov, Nucl. Phys. B 583 (1995) 711; \\ M.~Ga\'zdzicki et al.,
  Z. Physik  C 33
(1986) 895.
\bibitem{P94}
 S.~Pratt, Phys. Rev. C 49 (1994) 2722.


\end{thebibliography}
\end{document}